\pdfoutput=1
\documentclass[a4paper,11pt]{article}
\usepackage{amsmath,amssymb,amsfonts}
\usepackage[multiple,stable]{footmisc}
\usepackage[amsmath,amsthm,thmmarks]{ntheorem}
\allowdisplaybreaks[4]
\usepackage[noconfig]{refstyle}
\usepackage[dvipsnames]{xcolor}
\usepackage[hyperfootnotes=false, linktocpage=true, colorlinks, citecolor=blue, linkcolor=blue, urlcolor=Maroon]{hyperref}
\usepackage{array,tabularx,booktabs,geometry,subfigure,graphicx,longtable}
\usepackage[bf]{caption}
\usepackage{tikz}
\usepackage{cite}
\usepackage{dsfont}
\usepackage{relsize}
\usetikzlibrary{shapes,arrows}
\tikzstyle{block} = [rectangle, draw, text width=7em, text centered, rounded corners, minimum height=3em]

\let\eqref=\relax
\newref{eq}{name={eq.~},Name={Eq.~},names={eqs.~},Names={Eqs.~},rngtxt={-},refcmd=(\ref{#1})}
\newref{f}{name={footnote~},Name={Footnote~},names={footnotes~},Names={Footnotes~}}
\newref{app}{name={appendix~},Name={Appendix~},names={appendixes~},Names={Appendixes~}}
\newref{tab}{name={table~},Name={Table~},names={tables~},Names={Tables~}}
\newref{ch}{name={chapter~},Name={Chapter~},names={chapters~},Names={Chapters~}}
\newref{sec}{name={section~},Name={Section~},names={sections~},Names={Sections~}}
\newref{fig}{name={figure~},Name={Figure~},names={figures~},Names={Figures~}}
\numberwithin{equation}{section}
\newcommand{\be}{\begin{equation}}
\newcommand{\ee}{\end{equation}}
\newcommand{\bea}{\begin{equation}\begin{aligned}}	% note: abbreviations for \begin{align} and \end{align} don't work!
\newcommand{\eea}{\end{aligned}\end{equation}}		% note: \begin{equation}\begin{split}... produces pdf/hyperref warnings:
\newcommand{\tr}{\mathrm{tr}}

\newcommand{\iddots}{\mathinner{\mkern2mu\raise1pt\hbox{.}\mkern2mu \raise4pt\hbox{.}\mkern2mu\raise7pt\hbox{.}\mkern1mu}}

\providecommand{\id}{\leavevmode\hbox{\small$\mathrm{1}$\kern-3.8pt\normalsize$\mathrm{1}$}}
\def\fnote#1#2{\begingroup\def\thefootnote{#1}\footnote{#2}
     \addtocounter{footnote}{-1}\endgroup}
\begin{document}

\vspace{1cm}

\title{
       {\Large \bf Vanishing Yukawa Couplings and \\ the Geometry of String Theory Models}}

\vspace{2cm}

\author{
Lara~B.~Anderson,${}^{1}$
James~Gray,${}^{1}$ \\
Magdalena Larfors,${}^{2,3}$ and
Matthew Magill,${}^{2}$ 
}
\date{}
\maketitle
\begin{center} {\small ${}^1${\it Department of Physics, 
Robeson Hall, Virginia Tech \\ Blacksburg, VA 24061, U.S.A.}\\[0.2cm]
       ${}^2${\it Department of Physics and Astronomy, Uppsala University,\\
       $~~~~~$ SE-751 20 Uppsala, Sweden.}\\[0.2cm]
       ${}^2${\it Centre for Particle Theory, Department of Mathematical Sciences,\\
       $~~~~~$ Durham University, South Road, Durham DH1 3LE, UK.}}\\

\fnote{}{lara.anderson@vt.edu}
\fnote{}{jamesgray@vt.edu}
\fnote{}{magdalena.larfors@physics.uu.se}
\fnote{}{matthew.magill@physics.uu.se}

\end{center}

\begin{abstract}
\noindent
We provide an overview of recent work which aims to understand patterns of vanishing Yukawa couplings that arise in models of particle physics derived from string theory. These patterns are seemingly linked to a plethora of different geometrical structures and our understanding of the subject has yet to be consolidated in a unified framework. This short note is based upon a talk that was given by one of the authors at the Nankai Symposium on Mathematical Dialogues. Therefore it is aimed at a mathematical audience of mixed academic background.
\end{abstract}

\thispagestyle{empty}
\setcounter{page}{0}
\newpage

%\tableofcontents

%%%%%%%%%%%%%%%%%%%%%%%%%%%%%%%%%%%%%%%%%%%%%%%%%%

\section{Introduction}

Yukawa couplings are parameters that appear in theories of physics which describe how strongly certain particles interact. In many string theory constructions these Yukawa couplings are found to vanish in a somewhat mysterious fashion \cite{Strominger:1985ks,Greene:1986bm,Greene:1986jb,Candelas:1987se,Distler:1987gg,Distler:1987ee,Greene:1987xh,Candelas:1990pi,Distler:1995bc,Braun:2006me,Bouchard:2006dn,Anderson:2010vdj,Anderson:2010tc,Buchbinder:2014sya,Blesneag:2015pvz,Blesneag:2016yag,Anderson:2021unr}. These vanishings are linked to interesting geometrical structures associated to those theories. In this brief note we review this topic, in a manner aimed at an audience of geometers of mixed academic background, and highlight some open questions in this setting. 

\vspace{0.2cm}

We will consider particle physics models derived from heterotic string theory. In detailing such a model two pieces of geometrical data must be specified. These are a Calabi-Yau threefold $X$ and a holomorphic poly-stable vector bundle of zero slope $V_X$ over that threefold base whose structure group is a subgroup of $E_8$ \cite{Green:1987mn}. Given this data, the particle content of the resulting physical theory, and the Yukawa couplings between those particles, can be computed in a quasi-topological fashion.

\vspace{0.1cm}

The particle content of a heterotic model is given by various cohomology groups of associated bundles to $V_X$ \cite{Green:1987mn}. As a simple example, in the case of an $SU(3)$ vector bundle, families of matter fields in the particle physics theory are in one to one correspondence with elements of  $H^1(V_X)$ and anti-families of matter fields with elements of $H^1(V^{\vee}_X)$. Superpotential Yukawa couplings are numbers describing how three of the families (or anti-families) in such a theory interact. As such they are given by a map which takes the following form (in this simple setting) \cite{Distler:1987ee}.
\begin{eqnarray} \label{map1}
H^1(V_X) \otimes H^1(V_X) \otimes H^1(V_X) \to H^3(\wedge^3V_X) =H^3({\cal O})= \mathbb{C}
\end{eqnarray}
In the last two equalities in the above expression we have used that the bundle has $SU(3)$ structure group and that $h^3({\cal O})=1$ on a Calabi-Yau threefold. If these cohomology groups are represented in terms of bundle valued forms $\omega_i \in H^1(V_X)$ for $i=1,\ldots,h^1(V_X)$ then this map can be represented by the following integral \cite{Green:1987mn}.
\begin{eqnarray} \label{int1}
\int_X \tr(\omega_i \wedge \omega_j \wedge \omega_k) \wedge \Omega
\end{eqnarray}
Here the $\tr$ indicates taking the $SU(3)$ invariant combination and $\Omega$ is the nowhere vanishing holomorphic three-form on X. A similar, but somewhat more complex, discussion to that given here exists for the case of some other structure groups besides $SU(3)$ \cite{Green:1987mn,Distler:1987ee,Anderson:2010vdj}.

In building models of phenomenological particle physics, a generic guideline is that, unless there is a good reason for a Yukawa coupling to vanish, then one should not set it to zero. After all, if a number could take any value in the complex plane it seems improbable that it would turn out to be precisely and exactly zero by chance. The canonical example of a `good reason' would be a symmetry that forbids the coupling from appearing in the theory. The surprise seen in string theory models (see \cite{Braun:2006me,Bouchard:2006dn,Blesneag:2015pvz,Blesneag:2016yag,Gray:2019tzn,Anderson:2021unr} for some examples) is that many of these couplings do vanish, even though there is no known associated symmetry that  would account for such a feature. There are several possible reasons that this phenomenon might occur. One is that there may be a, previously unknown, symmetry for such systems that has thus far been missed. If such a symmetry did exist it would be of great interest to the subject of string theory compactifications to understand it. Another possibility is that for these vanishings the `good reason' may be associated to quasi-topological properties of the map (\ref{map1}). This too would be exciting, in that it would provide a signature of the higher dimensional nature of the theory. From the point of view of the four-dimensional particle physics model these vanishings would be seemingly unexplained, and only the higher dimensional avatar would provide a reason as to why this effect had occurred\footnote{For particle physics experts, it should be noted that the couplings being discussed here are high energy superpotential parameters. If there is truly no symmetry involved in this phenomenon, then this effect would provide a seemingly unexplained set of vanishing which would appear in the initial conditions for a computation of running couplings. This could lead to unexplained textures at low energies}.		

\section{Vanishing Couplings and Geometrical Structures}

In this short note we will focus on just two types of effect that can lead to vanishing Yukawa couplings.

\begin{itemize}
\item Constraints descending from fibration structures of $X$
\item Constraints descending from the possible embeddings of $X$ into some larger ambient space
\end{itemize}

Let us start by considering the case where vanishing Yukawa couplings are caused by the existence of a torus fibration structure of $X$. A discussion of this situation can be found in \cite{Braun:2006me} and we summarize some of those results here. In such a case, we have a projection map from $X$ to the base of the fibration.
\begin{eqnarray}
X \stackrel{\pi}{\to} B_2 
\end{eqnarray}
For simplicity we will consider the case where $V_X$ pushes down to a bundle on $B_2$. In such a case we have that
\begin{eqnarray} \label{fibgrad}
H^1(X,V_X)= \bigoplus_{p,q}^{p+q=1} H^p(B_2,R^q\pi_* V_X)
\end{eqnarray}
This implies that we have a $(p,q)$ grading on $H^1(X,V_X)$, which we will denote by
\begin{eqnarray}
H^p(B_2, R^q \pi_*V_X) :=(p,q|V_X) \;.
\end{eqnarray}
This grading is preserved under mappings such as (\ref{map1}).
\begin{eqnarray} \label{gradmap}
(p_1,q_1|V_X) \otimes (p_2,q_2|V_X) \otimes (p_3,q_3|V_X) \to (p_1+p_2+p_3,q_1+q_2+q_3|\wedge^3 V_X)
\end{eqnarray}
The target in (\ref{map1}) is $(2,1)$ graded: $H^3(X,{\cal O}) = (2,1|{\cal O})$. Thus for a non-zero Yukawa coupling we require three matter fields with grading $(p_i,q_i)$ such that $p_1+p_2+p_3=2$ and $q_1+q_2+q_3=1$. In any other case the coupling will be zero. In the language of forms, via the condition (\ref{int1}), this is simply the statement that we require the form $\tr(\omega_i \wedge \omega_j \wedge \omega_k)$ to have two legs on the base and one on the fiber. This can then be paired with $\Omega$, which has a similar structure, except that the legs are holomorphic rather than anti-holomorphic, to form a top form which can be integrated to give a non-zero value of the coupling.

The above discussion is for the simple case where $V_X$ pushes down to a bundle on $B_2$. We have focussed on this situation in the interests of brevity in this short note. However, a somewhat modified expression analogous to (\ref{gradmap}) does exist in more general cases, as is discussed in \cite{Braun:2006me}. 

\begin{figure}[!h]\centering
\includegraphics[width=0.6\textwidth]{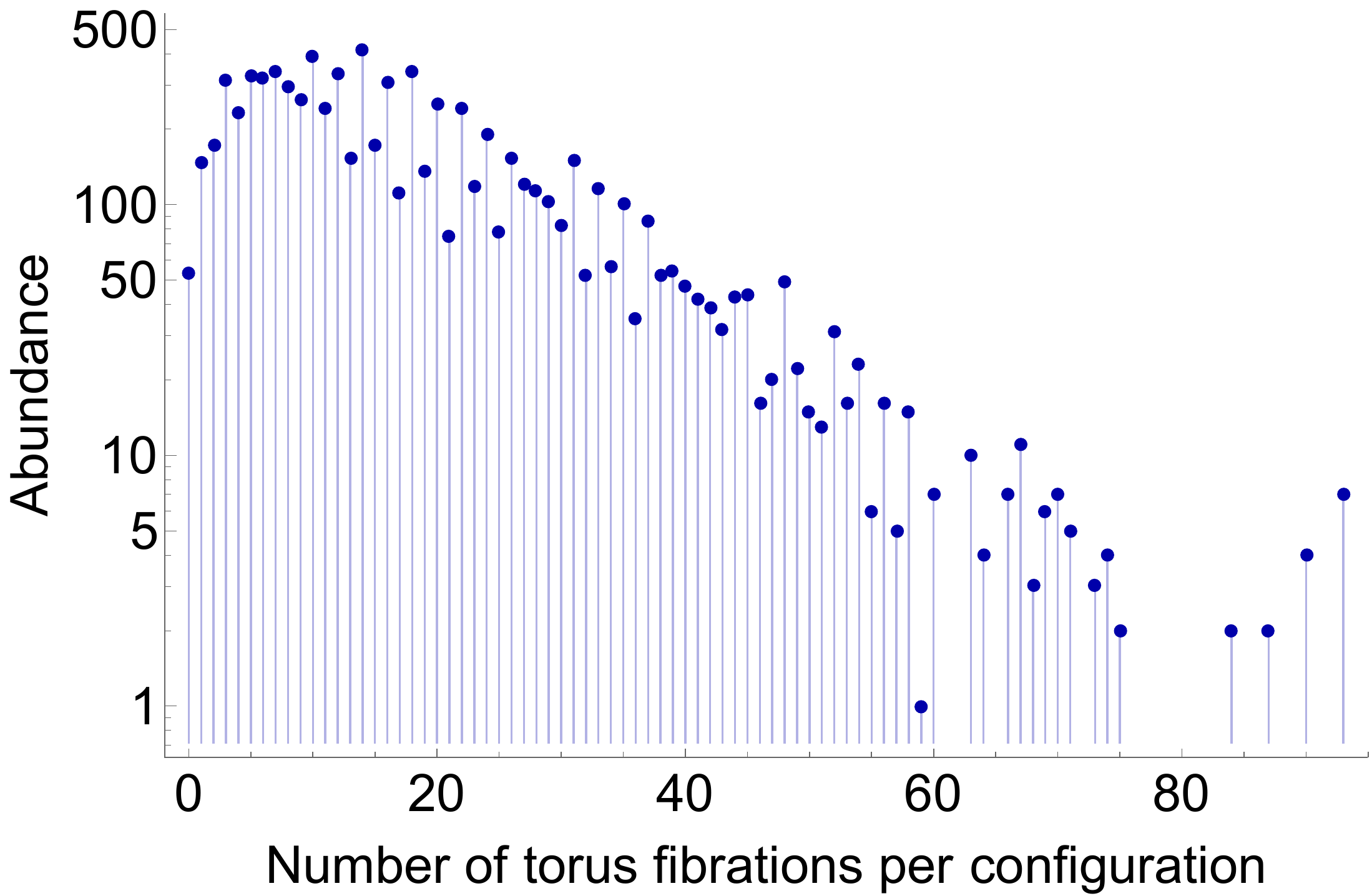}
\caption{{\it The number of Calabi-Yau threefolds (abundance), described as a complete intersection in products of projective spaces, that admit a given number of inequivalent torus fibrations. Figure taken from \cite{Anderson:2017aux}.}}
\label{fig1}
\end{figure}

The reason that constraints on Yukawa couplings due to fibrations are so important is that such substructures are ubiquitous in the known set of Calabi-Yau threefolds (see \cite{Gray:2014fla,Anderson:2016cdu,Anderson:2016ler,Anderson:2017aux,Anderson:2018kwv,Taylor:2012dr,Johnson:2014xpa,Huang:2018gpl,Huang:2018esr,Huang:2019pne} for studies of this in the physics literature and \cite{Filipazzi:2021dcw} for a recent discussion of this point in the mathematics literature). Indeed, in cases that have been analyzed in complete generality most Calabi-Yau manifolds are torus fibered in multiple different manners (see Figure \ref{fig1}) \cite{Rohsiepe:2005qg,Gray:2014fla,Anderson:2016cdu,Anderson:2017aux}.
Each of these different fibrations will lead to a different pattern of vanishing Yukawa couplings, all of which must then be manifested in the resulting physical theory.

\begin{figure}[!h]\centering
\includegraphics[width=0.6\textwidth]{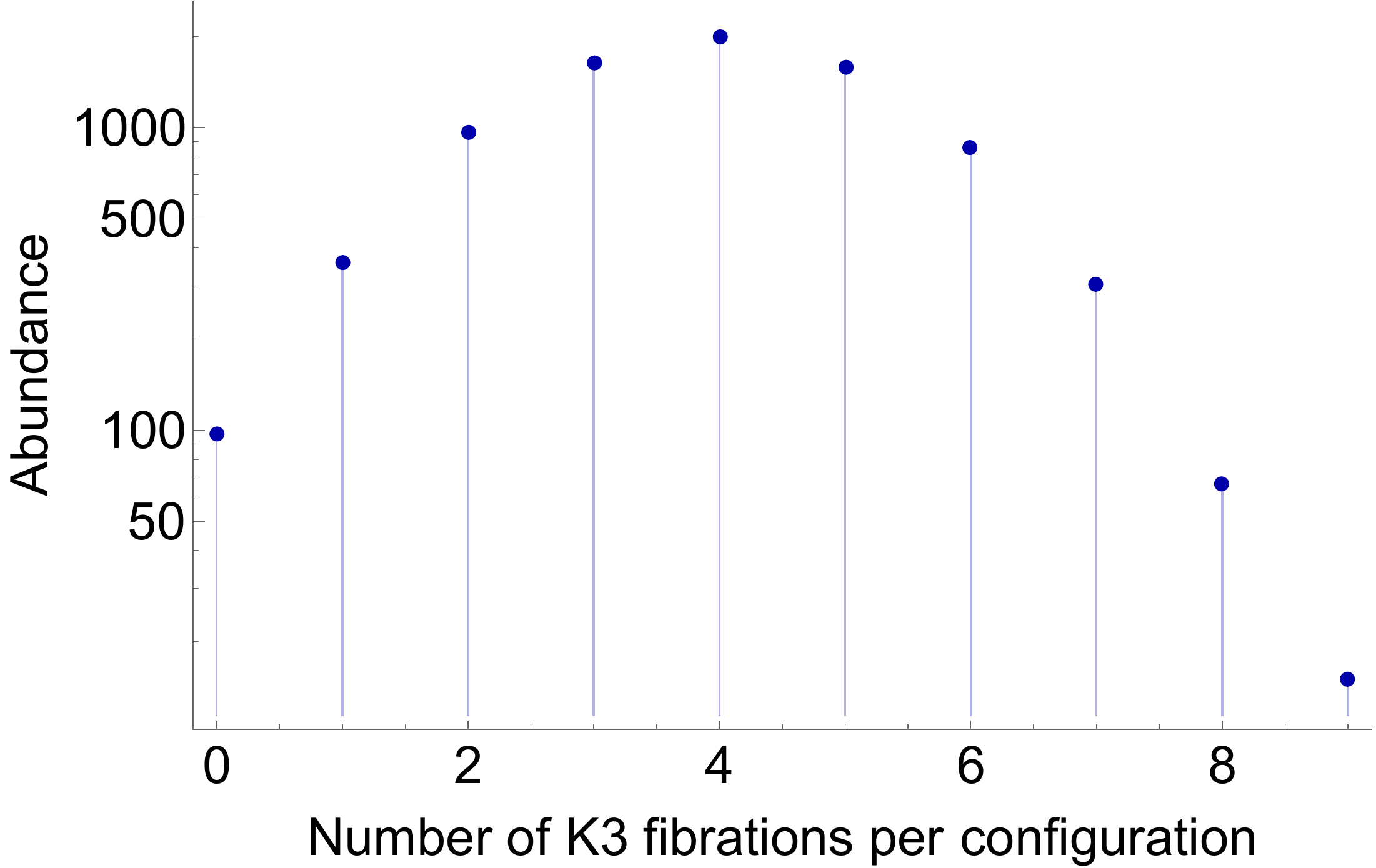}
\caption{{\it The number of Calabi-Yau threefolds (abundance), described as a complete intersection in products of projective spaces, that admit a given number of inequivalent K3 fibrations. Figure taken from \cite{Anderson:2017aux}.}}
\label{fig2}
\end{figure}

Note that one would expect other fibrations, such as $K3$ fibrations, to offer similar constraints, although to our knowledge this has not been systematically investigated in the literature to date. Such fibrations are also ubiquitous in the known set of Calabi-Yau threefolds, with most manifolds being multiply fibered by every smaller dimensional Calabi-Yau n-fold in cases that have been studied (see \cite{Anderson:2017aux} and Figure \ref{fig2}).

%%%%%%%%%%%%%%%%%%%%%%%%%%%%%%%%%%%%%%%%%%%%%%%%%%%
\vspace{0.3cm}
%%%%%%%%%%%%%%%%%%%%%%%%%%%%%%%%%%%%%%%%%%%%%%%%%%%	
	
Another source of vanishing Yukawa couplings stems from the description of Calabi-Yau manifolds as sub-loci inside a larger ambient space. The simplest form of this effect was described, and analyzed in a differential form associated to (\ref{int1}), in \cite{Blesneag:2015pvz,Blesneag:2016yag}. It was then analyzed in an algebraic language associated to (\ref{map1}) and generalized in \cite{Anderson:2021unr}.

To see the basic effect, consider a Calabi-Yau threefold $X$ embedded in an ambient space $A$. For simplicity we will consider $X$ to be defined as a hypersurface, with normal bundle ${\cal N}$, although the generalization to complete intersections is straight forward and included in the above references. Initially, we will also simplify the discussion to the case where the bundle $V_X$ on $X$ is the restriction of a bundle $V$ on the ambient space. The more general case where $V_X$ extends to a sheaf was considered in \cite{Anderson:2021unr} and we will return to that situation shortly. In the case where $V$ is a bundle we have a short exact Koszul sequence as follows.
\begin{eqnarray} \label{kos1}
0 \to {\cal N}^{\vee} \otimes V \to V \to V|_X \to 0
\end{eqnarray}
Studying the associated long exact sequence in cohomology, we find that the first cohomology of $V_X$ can be expressed in the following manner.
\begin{eqnarray} \label{ambgrad1}
H^1(V_X)=H^1(V|_X) = \text{coker} \left\{ H^1({\cal N}^{\vee} \otimes V) \to H^1(V) \right\} + \text{ker} \left\{ H^2({\cal N}^{\vee} \otimes V) \to H^2(V) \right\}
\end{eqnarray}
Now, as in (\ref{fibgrad}), we once again have a grading. Following \cite{Blesneag:2015pvz,Blesneag:2016yag}, a field is referred to as `type $i$' if it descends from the $i$'th ambient space cohomology. So for example, the fields associated to the $\text{coker}$ in (\ref{ambgrad1}) are of type 1 while those associated to the $\text{ker}$ are of type 2. The authors of \cite{Blesneag:2015pvz,Blesneag:2016yag} then go on to prove that the Yukawa couplings between three fields will vanish if the sum of their types is less than the dimension of the ambient space $A$.

This condition can already be restrictive when considering a description of $X$ in terms of just a single ambient space. For example, in \cite{Gray:2019tzn}, a study was performed of how many Yukawa couplings, allowed by every known symmetry, were forbidden by this effect in the largest data set of particle physics standard models that has been created using the construction under discussion \cite{Anderson:2011ns,Anderson:2012yf,Anderson:2013xka}. There it was shown that this effect lead to unexpected vanishings of Yukawa couplings in $30\%$  of the models in the data set.

The vanishings due to a single ambient space embedding of $X$, however, are not the full story in considering the implications of this phenomenon for Yukawa couplings in string theory models. Every Calabi-Yau manifold can be embedded in a huge variety of ambient spaces, and each of these descriptions of the manifold can potentially lead to different constraints on the Yukawa couplings. For example, the following three configuration matrices all describe the same Calabi-Yau threefold.
\begin{eqnarray}
&&\left[ \begin{array}{c|cccccccc}  \mathbb{P}^1&1&1 &0&0&0&0&0&0 \\ \mathbb{P}^1&0&0&1&0&0&0&1&0\\\mathbb{P}^1&1&0&0&1&0&0&0&0\\\mathbb{P}^1&0&0&0&0&1&1&0&0\\\mathbb{P}^1&1&0&0&0&0&0&0&1\\\mathbb{P}^2&0&1&0&0&0&0&1&1\\\mathbb{P}^2&0&0&1&0&1&0&0&1 \\\mathbb{P}^2&0&0&0&1&0&2&0&0 \end{array}\right] =\left[\begin{array}{c|ccccc} \mathbb{P}^1 &1&1&0&0&0 \\ \mathbb{P}^1 &0&0&2&0&0 \\ \mathbb{P}^1 & 0&0&0&0&2 \\ \mathbb{P}^1 &1&0&0&1&0\\ \mathbb{P}^1 &1&0&0&1&0 \\ \mathbb{P}^3 &0&1&1&1&1 \end{array} \right] \\ \nonumber &=& \left[ \begin{array}{c|cccccc}  \mathbb{P}^1 & 1&1&0&0&0&0 \\ \mathbb{P}^1 &0&0&1&1&0&0 \\ \mathbb{P}^1 & 0&0&0&0&2&0 \\ \mathbb{P}^1 &0&0&1&0&0&1\\ \mathbb{P}^1 &0&0&1&0&0&1\\ \mathbb{P}^4&1&1&0&1&1&1 \end{array} \right] = \left[ \begin{array}{c|ccccccccc} \mathbb{P}^1&0&0&0&1&1&0&0&0&0 \\\mathbb{P}^1&0&0&0&0&0&1&1&0&0 \\\mathbb{P}^1&1&0&0&0&0&0&0&1&0\\\mathbb{P}^1&0&1&0&0&0&0&0&0&1 \\\mathbb{P}^1&0&0&1&0&0&0&0&0&1\\\mathbb{P}^5&0&0&0&1&1&1&1&1&1 \\ \mathbb{P}^2&1&1&1&0&0&0&0&0&0 \end{array} \right]
\end{eqnarray}
In each of these matrices, the first column gives a series of projective spaces, the direct product of which is $A$. As we can see, in this simple example all of the ambient spaces are different. The remaining columns in each matrix each denotes the multi-degree of a defining polynomial of the variety with respect to the homogeneous coordinates of the projective ambient factors.
All of the different constraints on couplings descending from such different descriptions of $X$ must be realized in the physics of the compactification on that manifold, which should not depend on the description being used of the threefold.

To finish our discussion let us return briefly to the case where the bundle $V_X$ descends from a sheaf, not a bundle, on the ambient space $A$ in some description. In such a situation, the exact Koszul sequence (such as (\ref{kos1})) is replaced with a sequence which is does not terminate on the left. The above discussion can then be generalized by using the following result \cite{Anderson:2021unr}.

\vspace{0.2cm}

{\it Given any exact sequence,}
\begin{eqnarray}
 \ldots \to {\cal F}_2 \to {\cal F}_1 \to {\cal F}_0 \to V_X \to 0
\end{eqnarray}
{\it we can define a notion of `type' for elements of $H^1(V_X)$. Namely, we say that such a cohomology element has a type $\tau=i$ if it descends from an element of $H^i({\cal F}_{i-1})$. Then, if $H^3(\wedge^3 {\cal F}_0)=0$, all  Yukawa couplings between three type $\tau=1$ fields will vanish. If, in addition, $H^4({\cal F}_1\otimes \wedge^2 {\cal F}_0)=0$ then all Yukawa couplings between one type $\tau=2$ and two type $\tau=1$ fields will also vanish.}

\vspace{0.2cm}

This result was used in \cite{Anderson:2021unr} to show that, even for a given ambient space, different sheaf lifts of $V_X$ can lead to different constraints on the Yukawa couplings. As a simple example, consider the Calabi-Yau threefold given by the following configuration matrix.
\begin{eqnarray} \label{lastX}
X=\left[ \begin{array}{c|cc} \mathbb{P}^1 &1&1\\\mathbb{P}^1 &1&1\\\mathbb{P}^1 &1&1\\\mathbb{P}^1 &1&1\\\mathbb{P}^1 &1&1\\ \end{array} \right]
\end{eqnarray}
A simple sum of line bundles $V_X= {\cal O}_X(2,-2,0,0,0) \oplus {\cal O}_X(-1,1,0,0,0)^{\oplus 2}$ can descend from the ambient space of this example either from a bundle,
\begin{eqnarray} \label{bun1}
V_{\text{bundle}} &=& {\cal O}(2,-2,0,0,0) \oplus {\cal O}(-1,1,0,0,0)^{\oplus 2}\;,
\end{eqnarray}
or from a sheaf
\begin{eqnarray} \label{sheaf1}
V_{\text{sheaf}} &=& {\cal O}(2,-2,0,0,0) \otimes {\cal I}_L \oplus {\cal O}(-1,1,0,0,0)^{\oplus 2}\;.
\end{eqnarray}
In these expressions the arguments in the line bundles indicate the coefficients in an expansion of the first Chern class of that object in a basis of K\"ahler forms associated to the ambient space factors of (\ref{lastX}) (restricted to $X$ if necessary). In defining the ideal sheaf ${\cal I}_L$ we have defined a curve given by the following configuration matrix.
\begin{eqnarray} 
L = \left[ \begin{array}{c|cccc}  \mathbb{P}^1 &1&1&1&1 \\  \mathbb{P}^1 &1&1&1&1 \\  \mathbb{P}^1 &1&1&1&1 \\  \mathbb{P}^1 &1&1&1&1 \\  \mathbb{P}^1 &1&1&1&1 \\\end{array} \right]
\end{eqnarray}
The ideal sheaf itself is then defined via the usual short exact sequence.
\begin{eqnarray}
0\to {\cal I}_L \to {\cal O} \to {\cal O}_L \to 0
\end{eqnarray}

The constraints on Yukawa couplings descending from Koszul sequences for (\ref{bun1}) and (\ref{sheaf1}) are different. In particular, in this example, the sheaf lift of $V_X$ leads to no constraints on the Yukawa couplings at all, whereas the bundle lift leads to every coupling being forbidden. Thus, different lifts of the bundle can lead to different constraints on the Yukawa couplings, even in a single ambient space description of $X$.

\section{Conclusions}

In this short note we have seen a plethora of geometrical structures that can lead to patterns of vanishing Yukawa couplings in string theory models. Most known Calabi-Yau threefolds admit multiple fibration structures, each of which can lead to such features in the physical theory. In addition Calabi-Yau manifolds can be embedded in a wide variety of ambient spaces, and most of these embeddings will lead to a pattern of vanishing Yukawa couplings. Further, even for a given ambient embedding, different extensions of the gauge bundle to the ambient space can lead to different patterns of vanishing Yukawa couplings. This entire plethora of possible vanishing patterns of Yukawa couplings must {\it all} be realized in the associated particle physics theory. As such, it is common for Yukawa couplings to vanish in these constructions, even if there is no apparent symmetry that would cause these numbers to be zero. In fact, there are other effects that can also lead to vanishing Yukawa couplings that we have not had space to cover in this short note. For example, structures in the K\"ahler cone associated to the poly-stability of the vector bundle $V_X$ \cite{Sharpe:1998zu,Anderson:2009sw,Anderson:2009nt,Anderson:2010tc} can also lead to zero couplings \cite{Kuriyama:2008pv,Anderson:2010tc,Buchbinder:2014sya}.

In studying these phenomena one is left with the inescapable feeling that something is being missed. While it is true that all of these disparate causes do lead to vanishing couplings, one might expect that this phenomenon could be understood in a more unified manner. This could either be in terms of a symmetry that has thus far been missed in these constructions, or something more exotic. For example, perhaps there is some unified quasi-topological framework which encapsulates all of the possible information about vanishing couplings in a manner that doesn't refer to ancillary structures such as particular ambient spaces.

Whether there is a previously unknown symmetry of these string theory constructions, or a more exotic mechanism that predicts otherwise unexpected vanishings of Yukawa couplings in particle physics models coming from string theory, it seems that there is an exciting structure underlying these models which is waiting to be discovered.

\section*{Acknowledgements}
The work of LA and JG is supported by the NSF grant PHY-2014086. M.L., M.M. and R.S. are funded in part by Vetenskapsr\aa det, under grant number 2020-03230. The authors would like to thank the organizers of the Nankai Symposium on Mathematical Dialogues where the talk upon which this short note is based was presented.

%%%%%%%%%%%%%%%%%%%%%%%%%%%%%%%%%%%%%%%%%%%%%%%%%%%%%%%%%%%%%%%%%%%%%%%%%%%%%%%%%%%%%%%%%%%%%%%%

\end{document}